\documentclass[12pt]{article}
\usepackage{epsfig}
\begin{document}
\title{Testing the $\Lambda(1520)$ hyperon in-medium width in near-threshold proton--nucleus reactions}
\author{E. Ya. Paryev\\
{\it Institute for Nuclear Research, Russian Academy of Sciences,}\\
{\it Moscow 117312, Russia}}

\renewcommand{\today}{}
\maketitle

\begin{abstract}
   In the framework of the nuclear spectral function approach for incoherent primary proton--nucleon and
   secondary pion--nucleon production processes we study the inclusive $\Lambda(1520)$ hyperon production
   in the interaction of 2.83 GeV protons with nuclei. In particular, the A and momentum dependences of the
   absolute and relative $\Lambda(1520)$ hyperon yields are investigated in a two scenarios for
   its in-medium width. Our model calculations show that the pion--nucleon production channel contributes
   distinctly to the "low-momentum" $\Lambda(1520)$ creation both in light and heavy nuclei in the chosen
   kinematics and, hence, has to be taken into consideration on close examination of the dependences of the
   $\Lambda(1520)$ hyperon yields on the target mass number with the aim to get information on its width in
   the medium. They also demonstrate that both the A dependence of the relative $\Lambda(1520)$ hyperon
   production cross section and momentum dependence of the absolute $\Lambda(1520)$ hyperon yield at incident
   energy of interest are appreciably sensitive to the $\Lambda(1520)$ in-medium width, which means that these
   observables may be an important tool to determine the above width.
\end{abstract}

\newpage

\section*{1. Introduction}

\hspace{1.5cm} The investigation of in-medium properties of hadrons (mainly light vector mesons $\rho$, $\omega$, $\phi$) produced
in nuclei by heavy--ion and elementary--particle beams  has received considerable interest in recent years (see, for example, [1])
in the context of observation of partial restoration of chiral symmetry--the fundamental symmetry of QCD in the limit of massless
quarks--in the hot and dense nuclear matter. The another interesting case of resonance medium renormalization, recently found
theoretically, is that of the $\Lambda(1520)$ hyperon where the hadronic model [2, 3] predicts appreciable medium effects in cold
nuclear matter, e.g., an increase of the width of low-momentum $\Lambda(1520)$ hyperons at saturation density by a factor of five
compared to its vacuum value ($\Gamma_{\Lambda(1520)}=15.6$ MeV). Whereas the mass shift of the
$\Lambda(1520)$ hyperon in nuclear matter is found to be small (about 2\% of its free mass at normal nuclear matter density $\rho_0$) [2, 3]. The similar medium effects were predicted for the
$\phi$ meson [4--7]. The possibility to learn about its total in-medium width has been considered in [8--15].
As a measure for the in-medium broadening of the $\phi$ meson
the A dependence of its production cross section in nuclei in proton-- and photon--induced reactions has been employed
in these works. The A dependence is governed by the absorption of the $\phi$ meson flux in nuclear matter, which in turn is determined, in particular, by the phi in-medium width. Such method has been also recently adopted to study
the properties of the $\omega$ in finite nuclei in $\gamma$--induced reactions [16--19].

    Following the above method and using results of a model [2, 3] for the $\Lambda(1520)$ (or $\Lambda^*$)
    selfenergy in the nuclear medium as well as an eikonal approximation to account for the absorption of the
    outgoing $\Lambda(1520)$ resonance, the authors of [20] have calculated the A dependence of the ratio between
    the total $\Lambda(1520)$ production cross section in heavy nucleus and a light one ($^{12}$C) in
    ${\gamma}A$ and $pA$ collisions. The calculations were performed within the local Fermi sea model at energies accessible at present experimental facilities like SPring-8, ELSA and COSY.
    They have been done also with considering only the relevant elementary one-step $\Lambda(1520)$ production
    mechanisms and thus with neglecting in part for $pA$ reactions the $\Lambda(1520)$ creation in the
    two-step processes with an intermediate pions. These processes may contribute to the $\Lambda(1520)$
    production in near-threshold $pA$ interactions, much as the secondary pion-induced $\phi$ creation channels
    contribute [14, 15] to the $\phi$ production in such interactions. It is clear that, in order to get a deeper
    insight into the in-medium $\Lambda(1520)$ width in near-threshold $pA$ reactions, it is necessary to clarify
    the relative role of the primary proton--nucleon and secondary pion--nucleon $\Lambda(1520)$ production
    processes in these reactions--the main purpose of the present investigation. Moreover, from the experimental
    point of view it is also important to evaluate not only the relative, but also the absolute $\Lambda(1520)$
    production rates, which are lacking in [20], to elucidate the possibility of its real experimental observation
    in present facilities.

     In this paper we perform a detailed analysis of the $\Lambda(1520)$ production in $pA$ interactions at 2.83 GeV beam energy. This initial energy was used in the recent measurements of proton-induced $\phi$ production
   in nuclei at the ANKE-COSY facility [21]. We present the detailed predictions for the absolute and relative
   $\Lambda(1520)$ hyperon yields from these interactions
   obtained in the framework of a nuclear spectral function approach [14, 15] for an incoherent primary proton--nucleon ($pp \to pK^+{\Lambda(1520)}$, $pn \to nK^+{\Lambda(1520)}$, $pn \to pK^0{\Lambda(1520)}$) and secondary pion--nucleon
(${\pi}^+n \to K^+{\Lambda(1520)}$, ${\pi}^0p \to K^+{\Lambda(1520)}$, ${\pi}^0n \to K^0{\Lambda(1520)}$,
${\pi}^-p \to K^0{\Lambda(1520)}$) $\Lambda(1520)$ production processes in a two scenarios for its in-medium width.
These predictions can be used as an important tool for possible extracting the valuable information on the
 $\Lambda(1520)$ in-medium width, for instance, from the data sample collected in the recent ANKE experiment [21]
 or from the data which could be taken in a devoted experiment at the CSR facility at Lanzhou (China).

\section*{2. The model and inputs}

\section*{2.1. Direct $\Lambda(1520)$ production mechanism}

\hspace{1.5cm} An incident proton can produce a $\Lambda(1520)$ resonance directly in the first inelastic
$pN$ collision due to the nucleon's Fermi motion.
Since we are interested in the near-threshold energy region, we have taken into account the following elementary processes which have the lowest free production threshold ($\approx$ 2.77 GeV):
\begin{equation}
p+p \to p+{K^+}+\Lambda(1520),
\end{equation}
\begin{equation}
p+n \to n+{K^+}+\Lambda(1520),
\end{equation}
\begin{equation}
p+n \to p+{K^0}+\Lambda(1520).
\end{equation}
Because the ${\Lambda(1520)}N$ elastic scattering, similarly to the $NN$ elastic collision, is expected to be
sharply anisotropic at the relevant high (see below) $\Lambda(1520)$ laboratory momenta, we will neglect in the
present study the loss of $\Lambda(1520)$ events in the given lab solid angle in the nuclear production due to
the $\Lambda(1520)$ quasielastic rescatterings in the target nucleus.
Moreover, since the $\Lambda(1520)$ hyperon pole mass in the medium is approximately not affected by medium
effects [2, 3] as well as for reason of reducing the possible uncertainty of our calculations due to the use in them the model
nucleon [22, 23] and kaon [23--25] selfenergies, we will also ignore the medium modification of the outgoing hadron masses in the present work.
Then, taking into consideration the $\Lambda(1520)$ resonance final-state absorption as well as assuming that the
$\Lambda(1520)$ hyperon--as a narrow
resonance--is produced and propagated with its pole mass $M_{\Lambda(1520)}$ (or $M_{\Lambda^*}$) at small laboratory angles of our main interest
\footnote{A choice of these angles has been motivated by the fact that in the threshold energy region $\Lambda(1520)$
hyperons are mainly emitted in forward directions.}
 and using
the results given in [14, 15], we can represent the inclusive cross section for the production on nuclei
$\Lambda(1520)$ hyperons with
the momentum ${\bf p}_{\Lambda(1520)}$ (or ${\bf p}_{\Lambda^*}$)
from the primary proton-induced reaction channels (1)--(3) as follows
\footnote{It is worth noting that for kinematical conditions of our interest the contribution to the $\Lambda(1520)$
production
on nuclei from the $\Lambda(1520)$ hyperons produced in primary $pN \to NK{\Lambda(1520)}$ channels (1)--(3) and decaying into the ${K^-}p$ pairs (whereby the $\Lambda(1520)$ can be detected, for example, by the ANKE spectrometer)
inside the target nucleus is found on the basis of the model developed in [26] to be negligible compared to the overall
$\Lambda(1520)$ yield from primary and secondary $\Lambda(1520)$ creation processes calculated for these conditions within the present approach. The latter yield corresponds to the $\Lambda(1520)$ decays outside the target nucleus.}
:
\begin{equation}
\frac{d\sigma_{pA\to {\Lambda(1520)}X}^{({\rm prim})}
({\bf p}_0)}
{d{\bf p}_{\Lambda^*}}=I_{V}[A]
\end{equation}
$$
\times
\left[\frac{Z}{A}\left<\frac{d\sigma_{pp\to pK^+{\Lambda(1520)}}({\bf p}_{0}^{'},M_{\Lambda^*},
{\bf p}_{\Lambda^*})}{d{\bf p}_{\Lambda^*}}\right>+
2\frac{N}{A}\left<\frac{d\sigma_{pn\to nK^+{\Lambda(1520)}}({\bf p}_{0}^{'},M_{\Lambda^*},{\bf p}_{\Lambda^*})}{d{\bf p}_{\Lambda^*}}\right>\right],
$$
where
\begin{equation}
I_{V}[A]=2{\pi}A\int\limits_{0}^{R}r_{\bot}dr_{\bot}
\int\limits_{-\sqrt{R^2-r_{\bot}^2}}^{\sqrt{R^2-r_{\bot}^2}}dz
\rho(\sqrt{r_{\bot}^2+z^2})
\end{equation}
$$
\times
\exp{\left[-\sigma_{pN}^{{\rm in}}A\int\limits_{-\sqrt{R^2-r_{\bot}^2}}^{z}
\rho(\sqrt{r_{\bot}^2+x^2})dx
-\int\limits_{z}^{\sqrt{R^2-r_{\bot}^2}}\frac{dx}
{\lambda_{\Lambda^*}(\sqrt{r_{\bot}^2+x^2},M_{\Lambda^*})}\right]},
$$
\begin{equation}
\lambda_{\Lambda^*}({\bf r},M_{\Lambda^*})=\frac{p_{\Lambda^*}}{M_{\Lambda^*}
\Gamma_{{\rm tot}}({\bf r},M_{\Lambda^*})}
\end{equation}
and
\begin{equation}
\left<\frac{d\sigma_{pN\to NK^+{\Lambda(1520)}}({\bf p}_{0}^{'},M_{\Lambda^*},
{\bf p}_{\Lambda^*})}
{d{\bf p}_{\Lambda^*}}\right>=
\int\int
P({\bf p}_t,E)d{\bf p}_tdE
\left[\frac{d\sigma_{pN\to NK^+{\Lambda(1520)}}(\sqrt{s},M_{\Lambda^*},{\bf p}_{\Lambda^*})}
{d{\bf p}_{\Lambda^*}}\right],
\end{equation}
\begin{equation}
  s=(E_{0}^{'}+E_t)^2-({\bf p}_{0}^{'}+{\bf p}_t)^2,
\end{equation}
\begin{equation}
 E_{0}^{'}=E_{0}-\frac{{\Delta{\bf p}}^2}{2M_A},
\end{equation}
\begin{equation}
 {\bf p}_{0}^{'}={\bf p}_{0}-\Delta{\bf p},
\end{equation}
\begin{equation}
 \Delta{\bf p}=\frac{E_0V_0}{p_0}\frac{{\bf p}_{0}}{|{\bf p}_{0}|},
\end{equation}
\begin{equation}
   E_t=M_A-\sqrt{(-{\bf p}_t)^2+(M_{A}-m_{N}+E)^{2}}.
\end{equation}
Here,
$d\sigma_{pN\to NK^+{\Lambda(1520)}}(\sqrt{s},M_{\Lambda^*},{\bf p}_{\Lambda^*}) /d{\bf p}_{\Lambda^*}$
are the off-shell
differential cross sections for $\Lambda(1520)$ production in reactions (1) and (2)
at the $pN$ center-of-mass energy $\sqrt{s}$;
$\rho({\bf r})$ and $P({\bf p}_t,E)$ are the nucleon density and
nuclear spectral function normalized to unity;
${\bf p}_t$ and $E$ are the internal momentum and removal energy of the struck target nucleon
just before the collision; $\sigma_{pN}^{{\rm in}}$ and $\Gamma_{{\rm tot}}({\bf r},M_{\Lambda^*})$
are the inelastic cross section
\footnote{We use $\sigma_{pN}^{{\rm in}}=30$ mb for considered projectile proton energy of
$\epsilon_0=2.83$ GeV [14].}
of free $pN$ interaction and total $\Lambda(1520)$ width in its rest frame, taken at the point ${\bf r}$ inside the nucleus and at the pole of the resonance;
$Z$ and $N$ are the numbers of protons and neutrons in
the target nucleus ($A=N+Z$), $M_{A}$  and $R$ are its mass and radius;
$m_{N}$ is the bare nucleon mass; ${\bf p}_0$ and $E_0$
are the momentum and total energy of the initial proton; $V_0$ is the nuclear optical potential that
this proton feels in the interior of the nucleus ($V_0 \approx 40~{\rm MeV}$). When obtaining eq. (4) we
used the fact that the $\Lambda(1520)$ hyperon production cross sections in reactions (2) and (3) are the
same due to the isospin symmetry.
  The first term in it describes the contribution to the $\Lambda(1520)$ resonance production on nuclei from the primary $pp$ interaction (1), whereas the second one represents the contribution to this production from the $pn$ elementary processes (2) and (3). The quantity $I_V[A]$ in eq. (4) represents the effective number of target
  nucleons participating in the primary $pN \to NK\Lambda(1520)$ reactions. It is calculated according to
  equation (5) in which the first and the second exponential factors describe, respectively, the distortion
  of the initial proton inside the nucleus till the reaction point as well as the attenuation of the outgoing
 $\Lambda(1520)$ hyperon in its way out of the nucleus.

   Let us now specify the off-shell differential cross sections
$d\sigma_{pN\to NK^+{\Lambda(1520)}}(\sqrt{s},M_{\Lambda^*},{\bf p}_{\Lambda^*}) /d{\bf p}_{\Lambda^*}$
for $\Lambda(1520)$ production in the reactions (1), (2), entering into eqs. (4), (7). Following refs. [14, 15, 26--28], we assume that these
cross sections are equivalent to the respective on-shell cross sections calculated for the off-shell kinematics of the elementary processes  (1), (2).
In our approach the differential cross sections for $\Lambda(1520)$ production in the reactions (1), (2) have been
described by the three-body phase space calculations normalized to the corresponding total cross sections
$\sigma_{pN \to NK^+{\Lambda(1520)}}(\sqrt{s})$ [14, 15]:
\begin{equation}
\frac{d\sigma_{pN\rightarrow NK^+{\Lambda(1520)}}(\sqrt{s},M_{\Lambda^*},{\bf p}_{\Lambda^*})}
{d{\bf p}_{\Lambda^*}}
={\frac{{\pi}}{4E_{\Lambda^*}}}
{\frac{\sigma_{pN\rightarrow NK^+{\Lambda(1520)}}({\sqrt{s}})}
{I_{3}(s,m_N,m_K,M_{\Lambda^*})}}
{\frac{\lambda(s_{NK},m_{N}^{2},m_{K}^{2})}{s_{NK}}},
\end{equation}
\begin{equation}
I_{3}(s,m_N,m_{K},M_{\Lambda^*})=(\frac{{\pi}}{2})^2
\int\limits_{(m_{N}+m_{K})^2}^{({\sqrt{s}}-M_{\Lambda^*})^2}
{\frac{\lambda(s_{NK},m_{N}^{2},m_{K}^{2})}{s_{NK}}}
{\frac{\lambda(s,s_{NK},M_{\Lambda^*}^{2})}{s}\,ds_{NK}},
\end{equation}
\begin{equation}
\lambda(x,y,z)=\sqrt{{\left[x-({\sqrt{y}}+{\sqrt{z}})^2\right]}{\left[x-
({\sqrt{y}}-{\sqrt{z}})^2\right]}},
\end{equation}
\begin{equation}
s_{NK}=s+M_{\Lambda^*}^{2}-2(E_{0}^{'}+E_t)E_{\Lambda^*}+
2({\bf p}_{0}^{'}+{\bf p}_t){\bf p}_{\Lambda^*}.
\end{equation}
Here, $E_{\Lambda^*}$ is the total energy of a $\Lambda(1520)$ hyperon ($E_{\Lambda^*}=\sqrt{p_{\Lambda^*}^2+M_{\Lambda^*}^2}$) and $m_K$ is the bare kaon mass.

    Up to now, there have been no data on $\Lambda(1520)$ production in reactions (1) and (2).
    There is [29] only one data point (6.09$\pm$1.74)$~{\rm {\mu}b}$ for the free total cross section
$\sigma_{pn \to pK^0{\Lambda(1520)}}(\sqrt{s})$ of the reaction (3) at 0.8 GeV excess energy
\footnote{Which is defined as $\sqrt{s}-\sqrt{s_{th}}$, where $\sqrt{s_{th}}=m_p+m_K+M_{\Lambda^*}$
is the threshold energy.} (or at beam momentum of 6.5 GeV/c).
To estimate the $pN\to NK^+{\Lambda(1520)}$ free total cross sections
$\sigma_{pN \to NK^+{\Lambda(1520)}}(\sqrt{s})$, entering into eq. (13), we have employed the one-pion-exchange
model [30--32]. This model reasonably reproduces the available experimental data from many nucleon--nucleon
collision processes in a few GeV energy region. The one-pion-exchange diagram, for example, for the reaction
$pp\to pK^+{\Lambda(1520)}$ is shown in figure 1.
\begin{figure}[htb]
\begin{center}
\includegraphics[width=8.0cm]{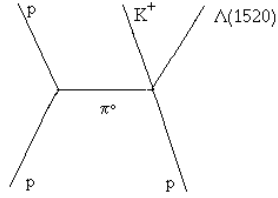}
\vspace*{-2mm} \caption{Diagram representing process (1).}
\label{void}
\end{center}
\end{figure}
Assuming the on-shell approximation for the virtual process
${\pi}^0p\to K^+{\Lambda(1520)}$, the total cross section for this reaction at the center-of-mass energy $\sqrt{s}$
can be written as [30--32]:
\begin{equation}
\sigma_{pp \to pK^+{\Lambda(1520)}}(\sqrt{s})=2\sigma_{A}(\sqrt{s}),
\end{equation}
\begin{equation}
\sigma_{A}(\sqrt{s})=\frac{1}{(2\pi)^2(\stackrel {*}{p_0}\sqrt{s})^2}\frac{m_p^2}{m_{\pi}^2}f_{{\pi}NN}^2
\end{equation}
$$
\times
\int\limits_{(m_{K}+{M_{\Lambda^*}})}^{({\sqrt{s}}-m_p)}k\,
s_1\,d(\sqrt{s_{1}})\sigma_{{\pi}^0p\to K^+{\Lambda(1520)}}(\sqrt{s_1})
\int\limits_{t_{-}}^{t_{+}}(-t)\,dt\,D^2(t)F^2(t),
$$
where
\begin{equation}
k=\frac{1}{2\sqrt{s_1}}\lambda(s_1,m_{p}^2,m_{\pi}^{2}),
\end{equation}
\begin{equation}
t_{\pm}=2m_p^2-2\stackrel {*}{E_0}\stackrel {*}{E_f}\pm~2\stackrel {*}{p_0}\stackrel {*}{p_f},
\end{equation}
\begin{equation}
\stackrel {*}{p_0}=\frac{1}{2\sqrt{s}}\lambda(s,m_{p}^2,m_{p}^{2}),\,\,\,\,\,\,\,\,
\stackrel {*}{p_f}=\frac{1}{2\sqrt{s}}\lambda(s,s_{1},m_{p}^{2}),
\end{equation}
\begin{equation}
\stackrel {*}{E_0}=\sqrt{\stackrel {*}{p_0}^2+m_p^2},\,\,\,\,\,\,\,\,\,\,
\stackrel {*}{E_f}=\sqrt{\stackrel {*}{p_f}^2+m_p^2}
\end{equation}
and
\begin{equation}
D(t)=\frac{1}{t-m_{\pi}^{2}},
\end{equation}
\begin{equation}
F(t)=\frac{\Lambda_{\pi}^2-m_{\pi}^2}{\Lambda_{\pi}^2-t}.
\end{equation}
Here, $m_{\pi}$ is the pion mass, $\Lambda_{\pi}$ is the respective cut-off parameter,
$f_{{\pi}NN}$ is the coupling constant
\footnote{Which was taken in the calculations to be equal to 1.0 [31, 32].}
,
and
$\sigma_{{\pi}^0p\to K^+{\Lambda(1520)}}(\sqrt{s_1})$ is the on-shell total cross section for the
reaction ${{\pi}^0p\to K^+{\Lambda(1520)}}$ at the ${\pi}N$ center-of-mass energy $\sqrt{s_1}$.
The quantity $\lambda$, entering into eqs. (19) and (21), is defined above by the formula (15).
According to the isospin considerations, we have:
\begin{equation}
\sigma_{{\pi}^0p\to K^+{\Lambda(1520)}}=\frac{1}{2}~\sigma_{{\pi}^-p\to K^0{\Lambda(1520)}}.
\end{equation}
For the free total cross section $\sigma_{{\pi}^-p\to K^0{\Lambda(1520)}}$ we have used the following
parametrization:
\begin{equation}
\sigma_{{\pi}^-p \to K^0{\Lambda(1520)}}({\sqrt{s_1}})=\left\{
\begin{array}{ll}
	123\left(\sqrt{s_1}-\sqrt{s_0}\right)^{0.47}~[{\rm {\mu}b}]
	&\mbox{for ${0}<\sqrt{s_1}-\sqrt{s_0}\le 0.427~{\rm GeV}$}, \\
	&\\
                   26.6/\left(\sqrt{s_1}-\sqrt{s_0}\right)^{1.33}~[{\rm {\mu}b}]
	&\mbox{for $\sqrt{s_1}-\sqrt{s_0} > 0.427~{\rm GeV}$},
\end{array}
\right.	
\end{equation}
where $\sqrt{s_0}=2.017$ GeV is the threshold energy for ${{\pi}^-p \to K^0{\Lambda(1520)}}$ reaction.
The comparison of the results of our calculations by (26) (solid line) with the available experimental data
(full squares) [33] for this reaction is shown in figure 2.
\begin{figure}[htb]
\begin{center}
\includegraphics[width=8.0cm]{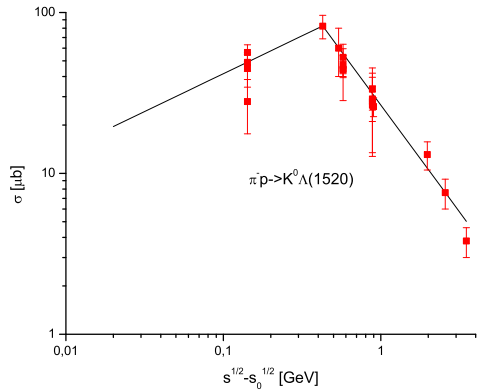}
\vspace*{-2mm} \caption{Total cross section for $\Lambda(1520)$ production in the reaction
${\pi^-}p \to K^0\Lambda(1520)$ as a function of excess energy $\sqrt{s}-\sqrt{s_0}$. For notation see the text.}
\label{centered}
\end{center}
\end{figure}
It is seen that our parametrization (26) fits well
the existing set of data for the ${{\pi}^-p \to K^0{\Lambda(1520)}}$ reaction. For the cut-off parameter
$\Lambda_{\pi}$, entering into eq. (24), we use $\Lambda_{\pi}=0.7$ GeV in order to describe (see figure 3)
by means of the relation
\begin{equation}
\sigma_{pn \to pK^0{\Lambda(1520)}}=2.5\sigma_{pp \to pK^+{\Lambda(1520)}},
\end{equation}
which the one-pion-exchange model of [30--32] gives, and eqs. (17)--(26) the $pn \to pK^0{\Lambda(1520)}$
data point at 0.8 GeV excess energy mentioned above. In figure 3 we show also the calculated cross section of
the reaction $pp \to pK^+{\Lambda(1520)}$, for which there are presently no experimental data. Looking at this
figure and keeping in mind that
$\sigma_{pn \to nK^+{\Lambda(1520)}}=\sigma_{pn \to pK^0{\Lambda(1520)}}$
due to the isospin symmetry, one can see that for the incoming proton energy of 2.83 GeV of our interest
(which corresponds to the excess energy of 20 MeV) the value of the total $\Lambda(1520)$ hyperon production
cross section in proton--proton and proton--neutron collisions is of the order of 2 and 10 nb, respectively.
The latter values are very small, but one should expect to measure theirs on ANKE-at-COSY facility [34] in order
to check the results of our model calculations of the primary elementary total cross sections, which will be used
below to calculate the yield of $\Lambda(1520)$ hyperons from proton--nucleus interactions.
\begin{figure}[htb]
\begin{center}
\includegraphics[width=8.0cm]{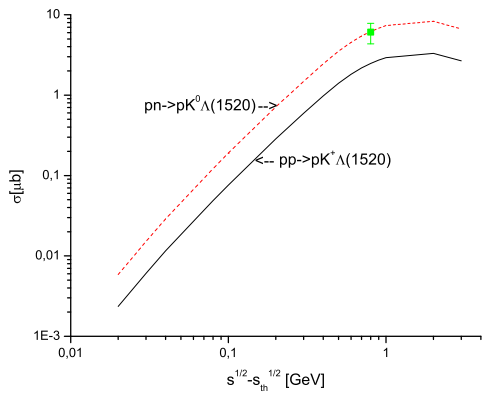}
\vspace*{-2mm} \caption{Total cross sections for the reactions $pp \to pK^+\Lambda(1520)$ (solid line) and
$pn \to pK^0\Lambda(1520)$ (dotted line) as functions of excess energy $\sqrt{s}-\sqrt{s_{th}}$ calculated,
respectively, by (17)--(26) and (17)--(27) with $\Lambda_{\pi}=0.7$ GeV. The experimental data point for the
$pn \to pK^0\Lambda(1520)$ process is taken from [29].}
\label{void}
\end{center}
\end{figure}

       For the $\Lambda(1520)$ production calculations in the case of  $^{12}$C and $^{27}$Al, $^{63}$Cu, $^{108}$Ag, $^{197}$Au, $^{238}$U target nuclei
reported here we have employed for the nuclear density $\rho({\bf r})$,
respectively, the harmonic oscillator and the Woods-Saxon distributions, given in [14].
The nuclear spectral function $P({\bf p}_t,E)$ (which represents the
 probability to find a nucleon with momentum ${\bf p}_t$ and removal energy $E$ in the nucleus) for these target nuclei was taken from [14, 15, 35--38].

   Let us concentrate now on the total $\Lambda(1520)$ in-medium width appearing in (6) and used in the subsequent calculations of $\Lambda(1520)$ resonance attenuation in $pA$ interactions.

   For this width we adopt two different scenarios: {\bf i)} no in-medium effects and, correspondingly, the scenario
   with the free $\Lambda(1520)$ width
 \footnote{Experimentally, the $\Lambda(1520)$ hyperons can be detected (for example, by the ANKE spectrometer)
 from their $\Lambda(1520) \to K^-p$ decay channel, having branching ratio $\approx$ 23\%. The $\Lambda(1520)$
 decays $\Lambda(1520) \to {\bar K^0}n,~{\Sigma}\pi,~{\Lambda}2\pi$, which exhaust $\approx$ 75\% of its free
 width, do not go into this detection channel and hence lead to the loss (or absorption) of $\Lambda(1520)$
 flux inside the nucleus with respect to this channel. Due to the strong final-state interactions of the
 $K^-$ and $p$ decay products, it does not contribute practically to the $\Lambda(1520)$ production on
 nuclei (cf. footnote 2). All the above means, that the free $\Lambda(1520)$ width induces a reduction of the
 number of the detectable $\Lambda(1520)$ hyperons and should be taken into account in the calculation
 of $\Lambda(1520)$ production on nuclei.}
   (dotted line in figure 4); {\bf ii)} the sum of
   the free $\Lambda(1520)$ width and its collisional width of the type [2, 3, 20]
   55($\rho_N/\rho_0$) MeV, where $\rho_N$ is the nucleon density (solid line
   \footnote{It shows that in this scenario the resulting total (decay + collisional) width of the $\Lambda(1520)$
   hyperon grows as a function of the density and reaches rhe value of around 70 MeV at the density $\rho_0$,
   which by a factor of $\sim$ 5 bigger than the free one.}
   in figure 4).
\begin{figure}[htb]
\begin{center}
\includegraphics[width=8.0cm]{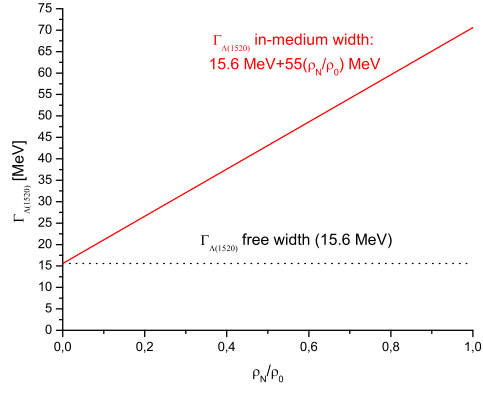}
\vspace*{-2mm} \caption{$\Lambda(1520)$ hyperon total width as a function of density. For notation see the text.}
\label{void}
\end{center}
\end{figure}
   According to [2, 3], the above collisional width is mainly governed by the following
   $\Lambda(1520)N$ interactions: $\Lambda(1520)N \to N\Lambda(1116)$, $\Lambda(1520)N \to N\Sigma(1189)$,
$\Lambda(1520)N \to N\Sigma(1385)$, $\Lambda(1520)N \to {\Delta}\Sigma(1189)$,
$\Lambda(1520)N \to {\Delta}\Sigma(1385)$. Therefore, the study of the $\Lambda(1520)$ absorption in nuclear
medium can help using the low-density approximation [1] to determine the unknown $\Lambda(1520)N$ inelastic cross
section. It should be noted that the adoption of the $\Lambda(1520)$ in-medium width, depicted by solid line
in figure 4, is justified only for small $\Lambda(1520)$ laboratory momenta [2, 3]. For high $\Lambda(1520)$
momenta it is needed, strictly speaking, to account for in the calculations the momentum dependence of this
width which emerges in particular from the (moderate) momentum dependence of the imaginary part,
$Im\Sigma_{\Lambda^*}$, of the $\Lambda(1520)$ selfenergy in nuclear matter found in [3] through the
relation [20]: $\Gamma_{\rm tot}=-(2E_{\Lambda^*}Im\Sigma_{\Lambda^*})/M_{\Lambda^*}$. However, because the
main goal of the present study was to clarify the role of the secondary pion--nucleon $\Lambda(1520)$
production process in near-threshold $pA$ reactions as well as for the sake of numerical simplicity, we neglect
the momentum dependence of the in-medium $\Lambda(1520)$ width $\Gamma_{\rm tot}$ in our calculations.
Evidently, this enables us to obtain an upper estimate of the strength of the respective double differential
cross sections and leads to the enlarged A dependence of the ratio
between the $\Lambda(1520)$ production cross section in heavy nucleus and a light one
especially in the region of heavy target nuclei where the absorption effect is strong (see below).

   Let us consider now the two-step $\Lambda(1520)$ production mechanism.

\section*{2.2. Two-step $\Lambda(1520)$ production mechanism}

\hspace{1.5cm} At the bombarding energy of our interest (2.83 GeV) the following two-step $\Lambda(1520)$ production process with a pion
in an intermediate state may contribute to the $\Lambda(1520)$ creation in $pA$ interactions:
\begin{equation}
p+N_1 \to \pi+X,
\end{equation}
\begin{equation}
\pi+N_2 \to K+\Lambda(1520),
\end{equation}
provided that the latter subprocess is allowed energetically
\footnote{We remind that the free threshold energy for this subprocess amounts to 1.55 GeV.}
.
Here, K stands for $K^+$ or $K^0$ for the specific isospin channel.
Taking into account the $\Lambda(1520)$ final-state absorption and ignoring the influence
\footnote{In line with the assumption about the absence of such influence on the final hadron masses in primary proton-induced
reaction channels (1)--(3).}
of the nuclear environment
on the outgoing hadron masses in the $\Lambda(1520)$ production channel (29) as well as using the results given in [14, 15], we get the
following expression for the $\Lambda(1520)$ production cross section for $pA$ reactions at small laboratory angles of interest from this
channel:
\begin{equation}
\frac{d\sigma_{pA\to {\Lambda(1520)}X}^{({\rm sec})}
({\bf p}_0)}
{d{\bf p}_{\Lambda^*}}=\frac{I_{V}^{({\rm sec})}[A]}{I_{V}^{'}[A]}
\sum_{\pi=\pi^+,\pi^0,\pi^-}\int \limits_{4\pi}d{\bf \Omega}_{\pi}
\int \limits_{p_{\pi}^{{\rm abs}}}^{p_{\pi}^{{\rm lim}}
(\vartheta_{\pi})}p_{\pi}^{2}
dp_{\pi}
\frac{d\sigma_{pA\to {\pi}X}^{({\rm prim})}({\bf p}_0)}{d{\bf p}_{\pi}}\times
\end{equation}
$$
\times
\left[\frac{Z}{A}\left<\frac{d\sigma_{{\pi}p\to K\Lambda(1520)}({\bf p}_{\pi},
{\bf p}_{\Lambda^*})}{d{\bf p}_{\Lambda^*}}\right>+\frac{N}{A}\left<\frac{d\sigma_{{\pi}n\to K\Lambda(1520)}({\bf p}_{\pi},
{\bf p}_{\Lambda^*})}{d{\bf p}_{\Lambda^*}}\right>\right],
$$
where
\begin{equation}
I_{V}^{({\rm sec})}[A]=2{\pi}A^2\int\limits_{0}^{R}r_{\bot}dr_{\bot}
\int\limits_{-\sqrt{R^2-r_{\bot}^2}}^{\sqrt{R^2-r_{\bot}^2}}dz
\rho(\sqrt{r_{\bot}^2+z^2})
\int\limits_{0}^{\sqrt{R^2-r_{\bot}^2}-z}dl
\rho(\sqrt{r_{\bot}^2+(z+l)^2})
\times
\end{equation}
$$
\times
\exp{\left[-\sigma_{pN}^{{\rm in}}A\int\limits_{-\sqrt{R^2-r_{\bot}^2}}^{z}
\rho(\sqrt{r_{\bot}^2+x^2})dx
-\sigma_{{\pi}N}^{{\rm tot}}A\int\limits_{z}^{z+l}
\rho(\sqrt{r_{\bot}^2+x^2})dx\right]}
\times
$$
$$
\times
\exp{\left[-\int\limits_{z+l}^{\sqrt{R^2-r_{\bot}^2}}\frac{dx}
{\lambda_{\Lambda^*}(\sqrt{r_{\bot}^2+x^2},M_{\Lambda^*})}\right]},
$$
\begin{equation}
I_{V}^{'}[A]=2{\pi}A\int\limits_{0}^{R}r_{\bot}dr_{\bot}
\int\limits_{-\sqrt{R^2-r_{\bot}^2}}^{\sqrt{R^2-r_{\bot}^2}}dz
\rho(\sqrt{r_{\bot}^2+z^2})\times
\end{equation}
$$
\times
\exp{\left[-\sigma_{pN}^{{\rm in}}A\int\limits_{-\sqrt{R^2-r_{\bot}^2}}^{z}
\rho(\sqrt{r_{\bot}^2+x^2})dx
-\sigma_{{\pi}N}^{{\rm tot}}A\int\limits_{z}^{\sqrt{R^2-r_{\bot}^2}}
\rho(\sqrt{r_{\bot}^2+x^2})dx\right]},
$$
\begin{equation}
\left<\frac{d\sigma_{{\pi}N\to K\Lambda(1520)}({\bf p}_{\pi},
{\bf p}_{\Lambda^*})}
{d{\bf p}_{\Lambda^*}}\right>=
\int\int
P({\bf p}_t,E)d{\bf p}_tdE
\left[\frac{d\sigma_{{\pi}N\to K\Lambda(1520)}(\sqrt{s_1},{\bf p}_{\Lambda^*})}
{d{\bf p}_{\Lambda^*}}\right];
\end{equation}
\begin{equation}
  s_1=(E_{\pi}+E_{t})^2-(p_{\pi}{\bf \Omega_{0}}+{\bf p}_{t})^2,
\end{equation}
\begin{equation}
 p_{\pi}^{{\rm lim}}(\vartheta_{\pi}) =
\frac{{\beta}_{A}p_{0}\cos{\vartheta_{\pi}}+
 (E_{0}+M_A)\sqrt{{\beta}_{A}^2-4m_{\pi}^{2}(s_{A}+
p_{0}^{2}\sin^{2}{\vartheta_{\pi}})}}{2(s_{A}+
p_{0}^{2}\sin^{2}{\vartheta_{\pi}})},
\end{equation}
\begin{equation}
 {\beta}_A=s_{A}+m_{\pi}^{2}-M_{A+1}^{2},\,\,s_A=(E_{0}+M_A)^2-p_{0}^{2},
\end{equation}
\begin{equation}
\cos{\vartheta_{\pi}}={\bf \Omega}_0{\bf \Omega}_{\pi},\,\,\,\,
{\bf \Omega}_{0}={\bf p}_{0}/p_{0},\,\,\,\,{\bf \Omega}_{\pi}={\bf p}_{\pi}/p_{\pi}.
\end{equation}
Here, $d\sigma_{pA\to {\pi}X}^{({\rm prim})}({\bf p}_0)/d{\bf p}_{\pi}$ are the
inclusive differential cross sections for pion production on nuclei at small laboratory angles and for high momenta from
the primary proton-induced reaction channel (28); $d\sigma_{{\pi}N\to K\Lambda(1520)}(\sqrt{s_1},{\bf p}_{\Lambda^*})/d{\bf p}_{\Lambda^*}$ is
the free inclusive differential cross section for $\Lambda(1520)$ production via the subprocess (29) calculated for the off-shell kinematics of
this subprocess at the ${\pi}N$ center-of-mass energy $\sqrt{s_1}$;
$\sigma_{\pi N}^{{\rm tot}}$ is the total cross section of the free $\pi N$ interaction
\footnote{We use in the following calculations $\sigma_{\pi N}^{{\rm tot}}=35$ mb for all pion momenta [14, 15].}
;
${\bf p}_{\pi}$ and $E_{\pi}$ are the momentum and total energy of a pion (which is assumed to be on-shell);
$p_{\pi}^{{\rm abs}}$ is the absolute threshold momentum for $\Lambda(1520)$ production on the residual nucleus by an intermediate pion;
$p_{\pi}^{{\rm lim}}(\vartheta_{\pi})$ is the kinematical limit for pion production
at the lab angle $\vartheta_{\pi}$ from proton-nucleus collisions. The quantity $\lambda_{\Lambda^*}$ is defined above by eq. (6). As is seen from eq. (30), we evaluate the $\Lambda(1520)$ yield for $pA$ reactions from the secondary
process (29) by folding the respective pion distribution from the primary proton-induced reaction channel (28)
(denoted by
\footnote{Where $I_{V}^{'}[A]$ is the effective number of target nucleons participating in this channel
(see eq. (32)).}
[$d\sigma_{pA\to {\pi}X}^{({\rm prim})}({\bf p}_0)/d{\bf p}_{\pi}$]/$I_{V}^{'}[A]$) with the
momentum--removal energy--averaged inclusive differential cross section for $\Lambda(1520)$ production
in this process (denoted by $\left<d\sigma_{{\pi}N\to K\Lambda(1520)}({\bf p}_{\pi},
{\bf p}_{\Lambda^*})/d{\bf p}_{\Lambda^*}\right>$) and with the effective number of $N_1N_2$ pairs per
unit of area (denoted via $I_{V}^{({\rm sec})}[A]$), involved in the two-step $\Lambda(1520)$ production
chain under consideration. In the expression (31), which defines the quantity $I_{V}^{({\rm sec})}[A]$,
the first two eikonal factors account for the distortion of the incident proton and intermediate pion
inside the target nucleus, whereas the latter one describes the attenuation of the produced $\Lambda(1520)$
hyperon in its way out of the nucleus.

    The expression for the differential cross section for $\Lambda(1520)$ production in the elementary process (29) has the following form:
\begin{equation}
\frac{d\sigma_{{\pi}N\rightarrow K\Lambda(1520)}(\sqrt{s_1},{\bf p}_{\Lambda^*})}
{d{\bf p}_{\Lambda^*}}
={\frac{{\pi}}{I_2(s_1,m_K,M_{\Lambda^*})E_{\Lambda^*}}}
{\frac{d\sigma_{{\pi}N\rightarrow K\Lambda(1520)}({\sqrt{s_1}})}
{d{\bf \Omega}_{\Lambda^*}^{*}}}\times
\end{equation}
$$
\times
{\frac{1}{(\omega+E_t)}}\delta\left[\omega+E_t-\sqrt{m_K^2+({\bf Q}+{\bf p}_t)^2}\right],
$$
where
\begin{equation}
I_2(s_1,m_K,M_{\Lambda^*})=\frac{\pi}{2}\frac{\lambda(s_1,m_{K}^{2},M_{\Lambda^*}^{2})}{s_1},
\end{equation}
\begin{equation}
\omega=E_{\pi}-E_{\Lambda^*}, \,\,\,\,{\bf Q}={\bf p}_{\pi}-{\bf p}_{\Lambda^*}.
\end{equation}
In eq. (38), $d\sigma_{{\pi}N\rightarrow K\Lambda(1520)}({\sqrt{s_1}})/d{\bf \Omega}_{\Lambda^*}^{*}$ is the
differential cross section for $\Lambda(1520)$ production in reaction (29) in the ${\pi}N$ c.m.s., which is assumed to be isotropic in our calculations of $\Lambda(1520)$ creation in $pA$ collisions from this reaction:
\begin{equation}
\frac{d\sigma_{{\pi}N\rightarrow K\Lambda(1520)}({\sqrt{s_1}})}
{d{\bf \Omega}_{\Lambda^*}^{*}}=\frac{\sigma_{{\pi}N\rightarrow K\Lambda(1520)}({\sqrt{s_1}})}{4\pi}.
\end{equation}
Here, $\sigma_{{\pi}N\rightarrow K\Lambda(1520)}({\sqrt{s_1}})$ is the total cross section of the elementary process ${\pi}N \to K\Lambda(1520)$.
The elementary $\Lambda(1520)$ production reactions  ${\pi}^+n \to K^+\Lambda(1520)$, ${\pi}^0p \to K^+\Lambda(1520)$, ${\pi}^0n \to K^0\Lambda(1520)$ and
${\pi}^-p \to K^0\Lambda(1520)$ have been included in our calculations of the $\Lambda(1520)$ production on nuclei.
The isospin considerations show that the following relations among the total cross sections of these reactions exist:
\begin{equation}
\sigma_{{\pi}^{+}n \to K^+\Lambda(1520)}=\sigma_{{\pi}^{-}p \to K^0\Lambda(1520)},
\end{equation}
\begin{equation}
\sigma_{{\pi}^{0}p \to K^+\Lambda(1520)}=\sigma_{{\pi}^{0}n \to K^0\Lambda(1520)}=\frac{1}{2}\sigma_{{\pi}^{-}p \to K^0\Lambda(1520)}.
\end{equation}
For the free total cross section $\sigma_{{\pi}^{-}p \to K^0\Lambda(1520)}$ we have used the suggested above
parametrization (26).

  Another very important ingredient for the calculation of the $\Lambda(1520)$ production cross section in proton--nucleus reactions
from pion-induced reaction channel (29)--the high-momentum parts of the differential cross sections for pion production on nuclei at small lab angles from the primary process (28) was taken from [14, 15, 37, 38].

  Now, let us proceed to the discussion of the results of our calculations for $\Lambda(1520)$ production in $pA$ interactions in the framework of the model outlined above.

\section*{3. Results and discussion}

\hspace{1.5cm} At first, we consider the absolute $\Lambda(1520)$ production cross sections from the one-step
and two-step $\Lambda(1520)$ creation mechanisms in $p^{12}$C, $p^{63}$Cu, $p^{197}$Au collisions calculated
on the basis of eqs. (4), (30) for proton kinetic energy of 2.83 GeV and for the laboratory $\Lambda(1520)$
production angle of 0$^{\circ}$ in two considered scenarios (see figure 4) for the total $\Lambda(1520)$
in-medium width. The cross sections calculated using the free $\Lambda(1520)$ width for this width are depicted
in figure 5.
\begin{figure}[htb]
{\centering
\includegraphics[width=5.0cm]{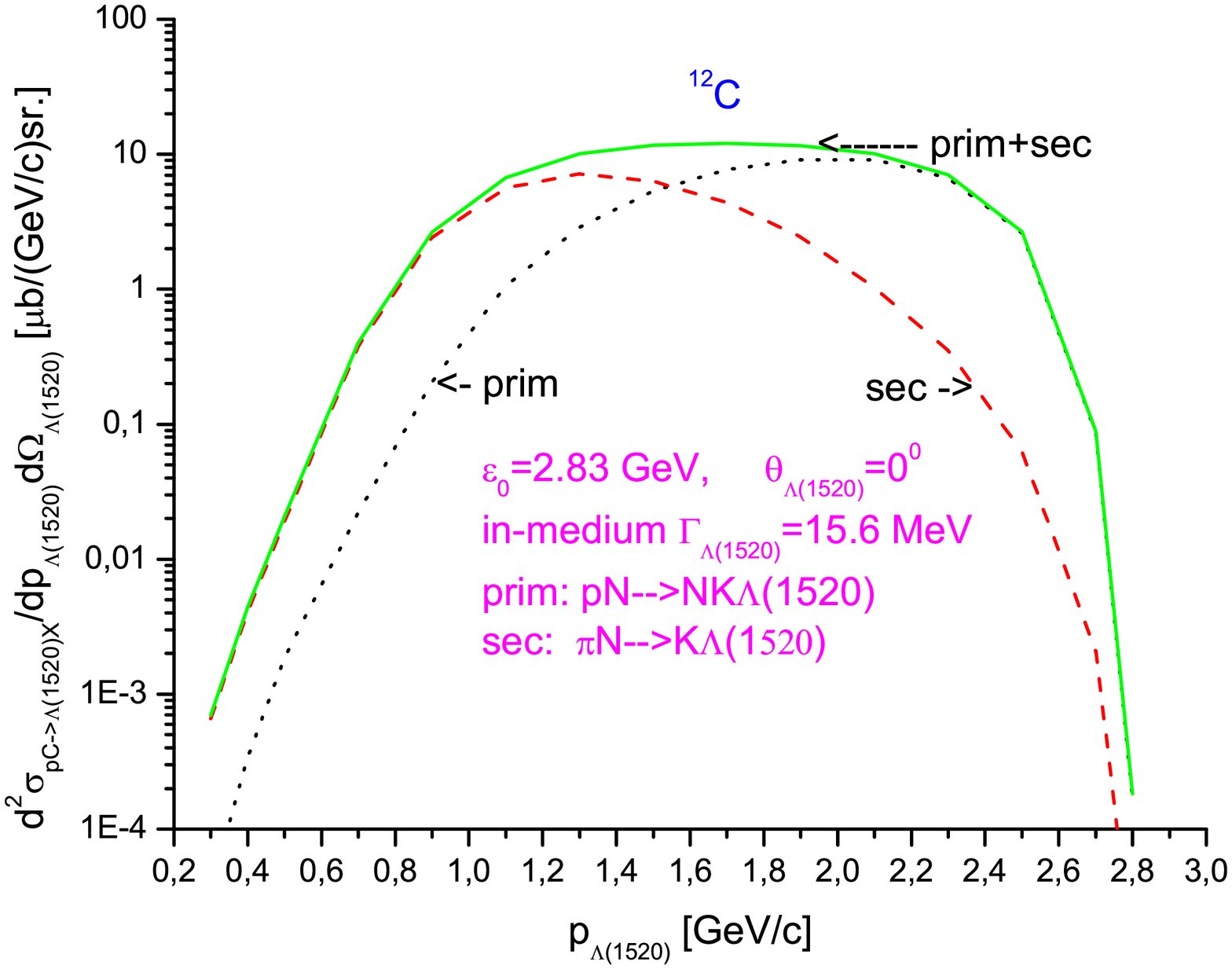}
\hfill
\includegraphics[width=5.0cm]{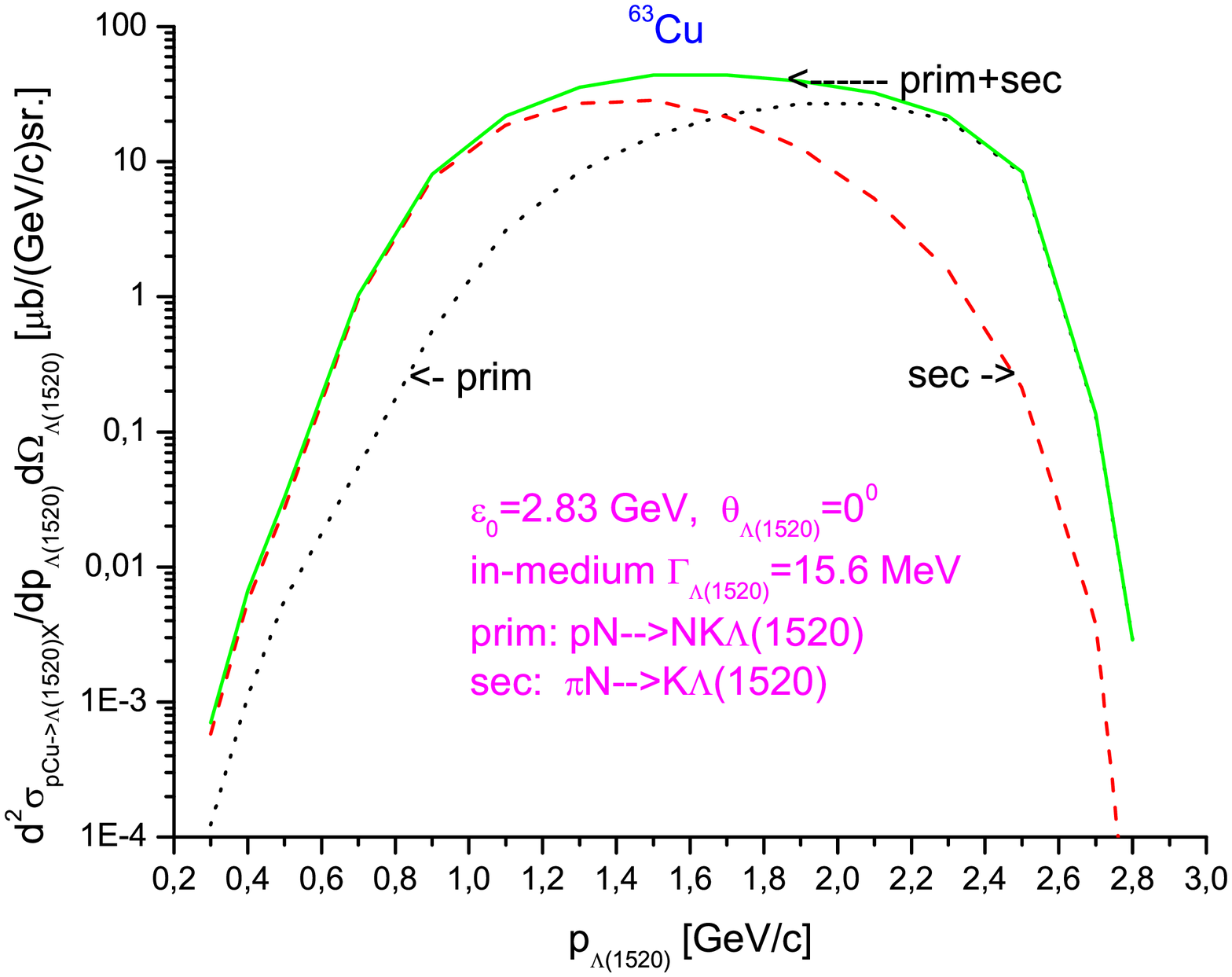}
\hfill
\includegraphics[width=5.0cm]{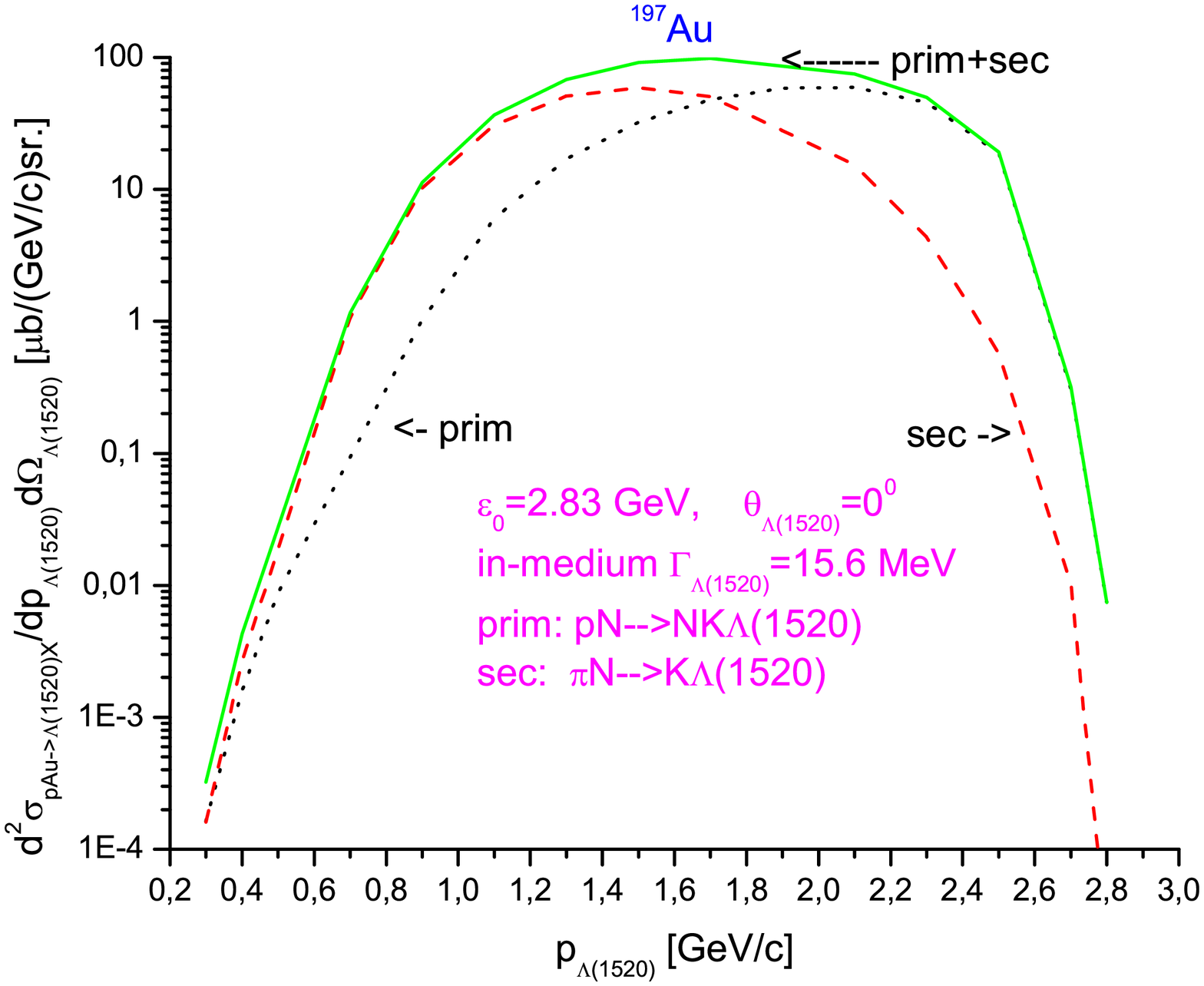}}
\vspace*{-2mm} \caption{Double differential cross sections for the production of $\Lambda(1520)$ hyperons
at a lab angle of 0$^{\circ}$ in the interaction of protons of energy 2.83 GeV with $^{12}$C (left panel),
$^{63}$Cu (middle panel) and $^{197}$Au (right panel) nuclei as
functions of $\Lambda(1520)$ momentum. The dotted and dashed lines are calculations, respectively, for the one-
and two-step $\Lambda(1520)$ creation mechanisms. The solid line is the sum of the dotted and dashed lines. The
loss of $\Lambda(1520)$ hyperons in nuclear matter was determined by their free width (dotted curve in figure 4).}
\label{void}
\end{figure}
Whereas the ones obtained for in-medium $\Lambda(1520)$ width shown by solid curve in figure 4
are given in figure 6.
\begin{figure}[htb]
{\centering
\includegraphics[width=5.0cm]{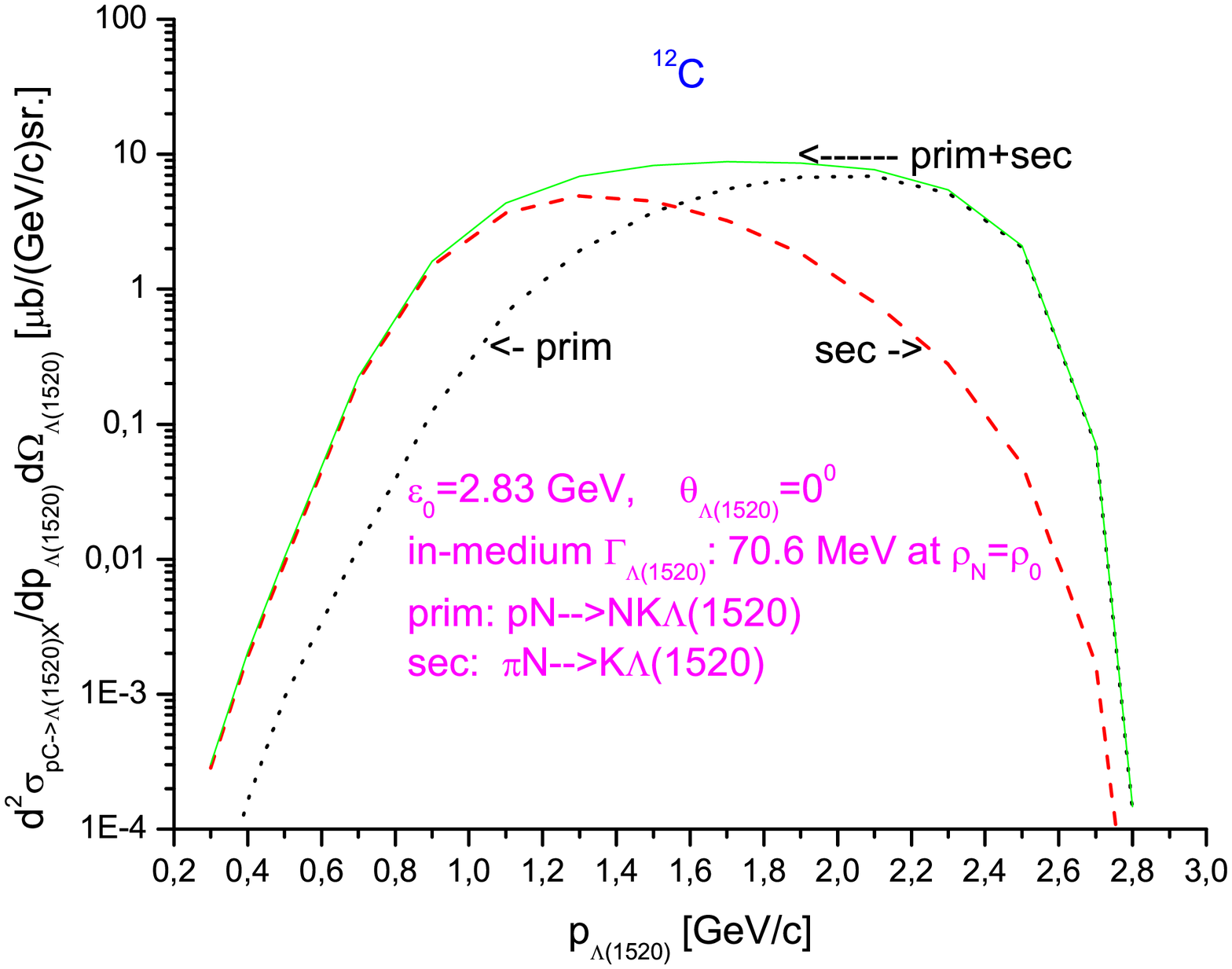}
\hfill
\includegraphics[width=5.0cm]{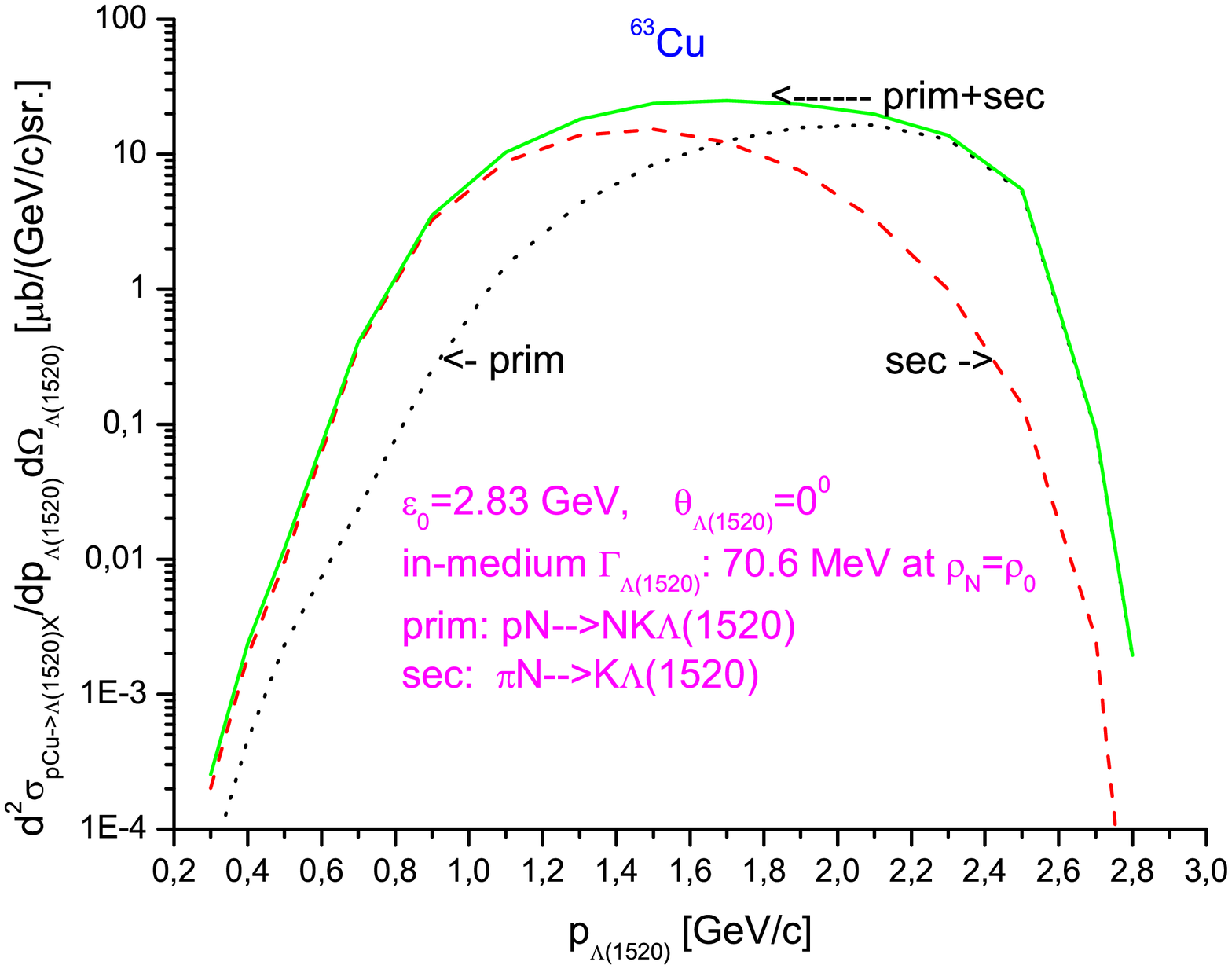}
\hfill
\includegraphics[width=5.0cm]{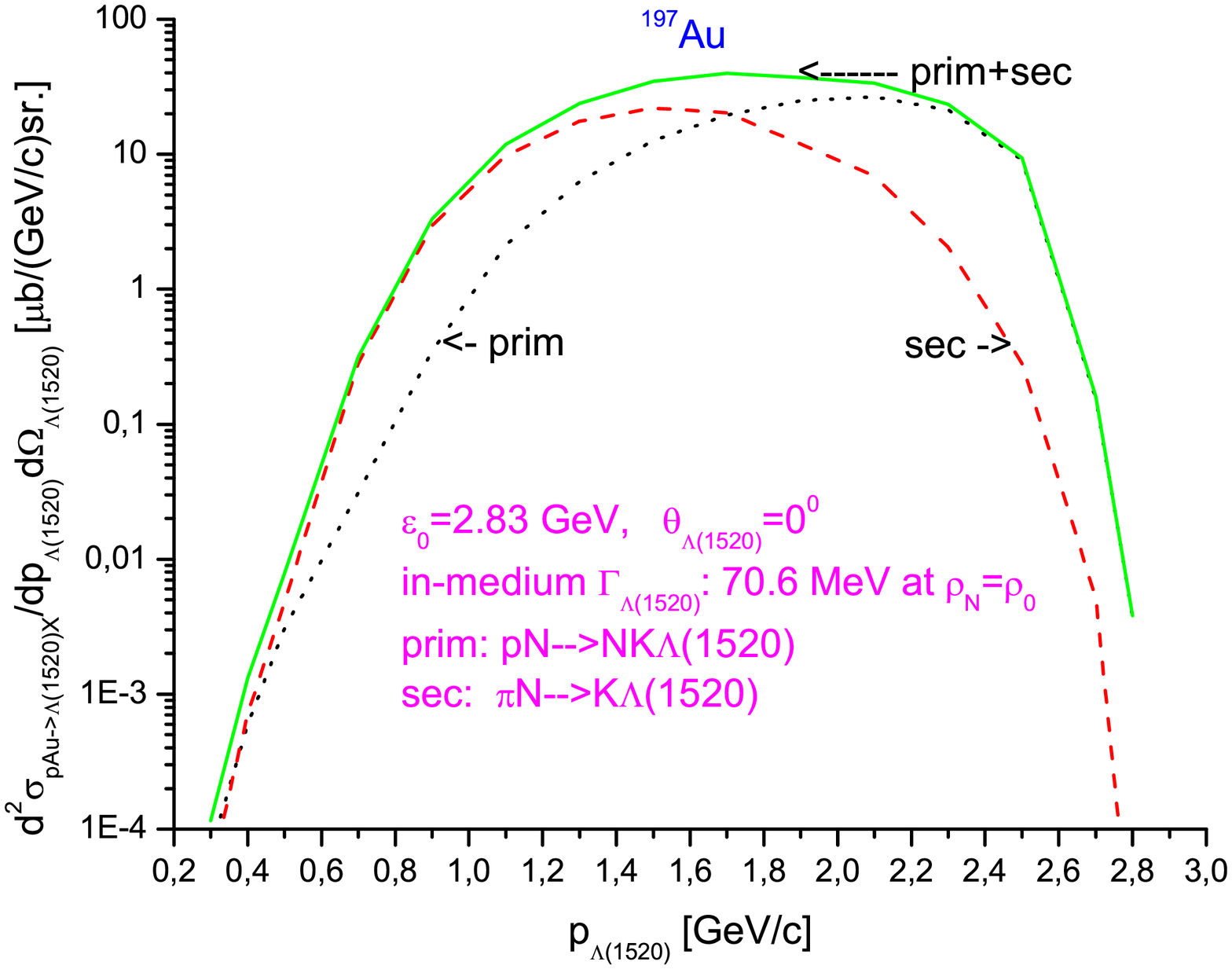}}
\vspace*{-2mm} \caption{Double differential cross sections for the production of $\Lambda(1520)$ hyperons
at a lab angle of 0$^{\circ}$ in the interaction of protons of energy 2.83 GeV with $^{12}$C (left panel),
$^{63}$Cu (middle panel) and $^{197}$Au (right panel) nuclei as
functions of $\Lambda(1520)$ momentum. The dotted and dashed lines are calculations, respectively, for the one-
and two-step $\Lambda(1520)$ creation mechanisms. The solid line is the sum of the dotted and dashed lines. The
absorption of $\Lambda(1520)$ hyperons in nuclear matter was governed by their in-medium width shown by
solid curve in figure 4.}
\label{void}
\end{figure}
Looking at these figures, one can see that the primary $pN \to NK\Lambda(1520)$ processes play the dominant role
at $\Lambda(1520)$ momenta $\ge$ 1.7 GeV/c for all considered target nuclei and for both adopted scenarios for the
$\Lambda(1520)$ in-medium width, whereas at lower $\Lambda(1520)$ momenta the secondary pion-induced reaction channel
${\pi}N \to K\Lambda(1520)$ is clearly dominant only for $^{12}$C and $^{63}$Cu target nuclei. In the case of having
$^{197}$Au as target nucleus, this channel dominates the $\Lambda(1520)$ production only at $\Lambda(1520)$
momenta $0.5~{\rm {GeV/c}} \le p_{\Lambda(1520)} \le 1.7~{\rm {GeV/c}}$, whereas at $\Lambda(1520)$ momenta
$\le$ 0.5 GeV/c its dominance, contrary to the results for $^{12}$C and $^{63}$Cu, is less pronounced. This means
that the channel ${\pi}N \to K\Lambda(1520)$ has to be taken into consideration on close examination
\footnote{In particular, of the A dependence of the ratio between the total $\Lambda(1520)$ production cross
section in heavy nucleus and a light one in $pA$ collisions. Let us remind that the latter observable has been
studied in [20] without accounting for the process ${\pi}N \to K\Lambda(1520)$.}
of the A dependence of the relative $\Lambda(1520)$ hyperon production cross section in proton--nucleus reactions
at energies just above threshold with the aim of extracting of the information on the $\Lambda(1520)$ width in
nuclear medium. Comparing the results of our full calculations (the sum of contributions both from primary and from
secondary $\Lambda(1520)$ production processes) presented in figures 5 and 6 by solid lines, we
see yet that for given target nucleus there is clear difference between the results obtained by using different
$\Lambda(1520)$ in-medium widths under consideration.
We may see, for example, that for the considered $^{12}$C, $^{63}$Cu and $^{197}$Au nuclei 
the cross section in the momentum range $\sim$ 1.2--2.4 GeV/c (where it is the greatest) 
when calculated with the increased $\Lambda(1520)$ width in the medium 
is reduced, respectively, by a factors of about 1.3, 1.6 and 2.3 compared to that obtained in the case
when the loss of $\Lambda(1520)$ hyperons in nuclear matter was determined by their free width.

     In figure 7 we show together the results of our overall calculations, given before separately in figures 5 and 6  by solid lines, for the double differential cross sections for the production of $\Lambda(1520)$ hyperons
     on $^{12}$C and $^{197}$Au target nuclei
     at a lab angle of 0$^{\circ}$ for the primary (1)--(3) plus secondary (29) $\Lambda(1520)$ creation processes
     obtained for bombarding energy of 2.83 GeV by employing in them both the free $\Lambda(1520)$ hyperons width
     (dotted lines) and their in-medium width shown by solid curve in figure 4 (solid lines) to see more clearly the
     sensitivity of the calculated cross sections to the choice of the $\Lambda(1520)$ in-medium width. It is nicely
     seen that, at least for heavy nuclei like $^{197}$Au, there are measurable changes in the absolute cross sections
     for $\Lambda(1520)$ production on these nuclei due to its in-medium width.
\begin{figure}[!h]
\begin{center}
\includegraphics[width=8.0cm]{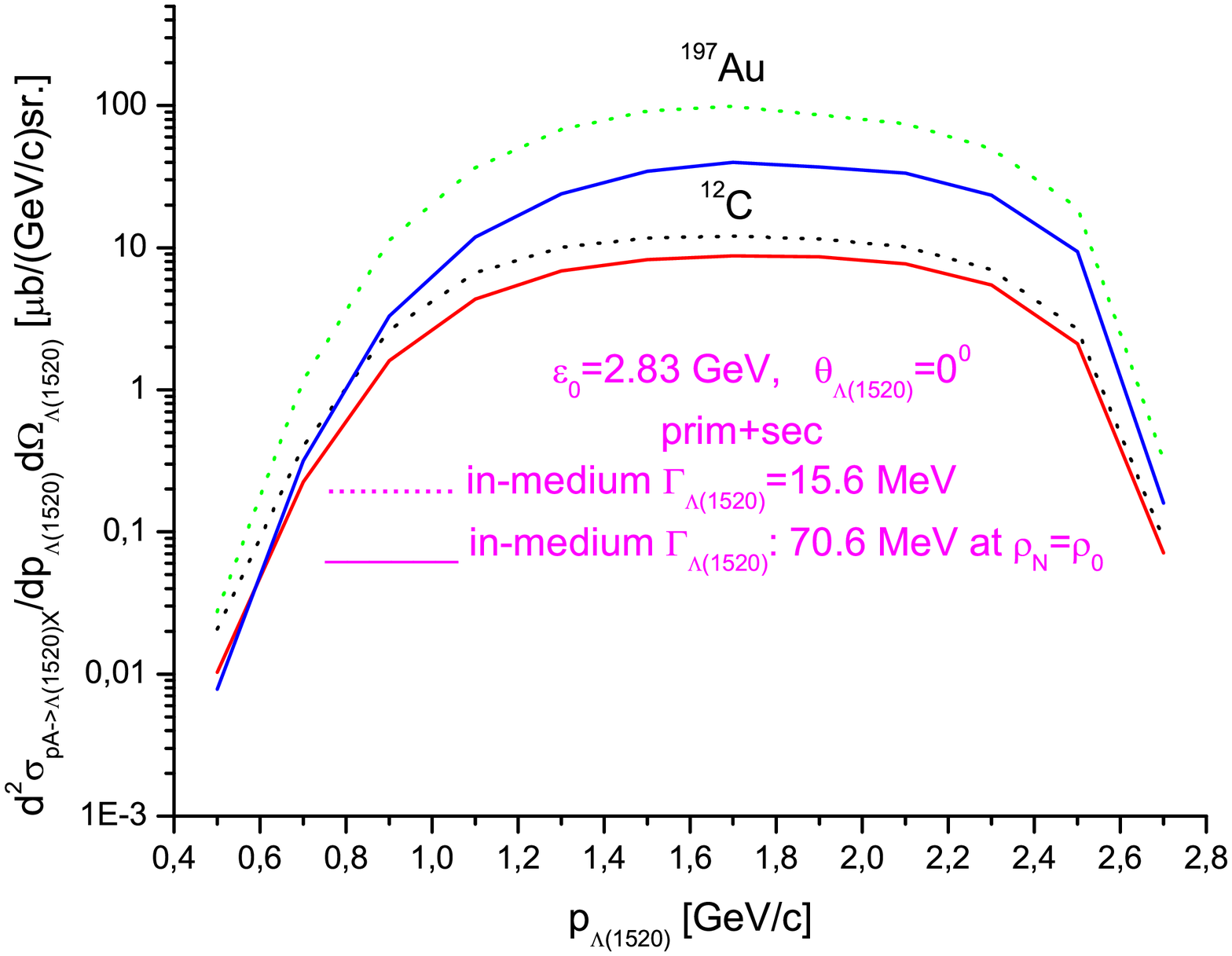}
\vspace*{-2mm} \caption{Double differential cross sections for the production of $\Lambda(1520)$ hyperons
at an angle of 0$^{\circ}$ in the interaction of protons of energy 2.83 GeV with $^{12}$C (two lower lines)
and $^{197}$Au (two upper lines) target nuclei
as functions of $\Lambda(1520)$ momentum for the one- plus two-step $\Lambda(1520)$
creation mechanisms calculated within the different scenarios for the $\Lambda(1520)$ in-medium width. The
dotted lines are calculations for the free $\Lambda(1520)$ width. The solid lines are calculations with
employing for the $\Lambda(1520)$ in-medium width that shown by solid curve in figure 4.}
\label{void}
\end{center}
\end{figure}

    We, therefore, come to the conclusion that the in-medium properties of the $\Lambda(1520)$ should be in principle observable through the
target mass dependence of the absolute $\Lambda(1520)$ production cross section in $pA$ reactions at above threshold beam energies.

    However, the authors of ref. [20] have suggested to use as a measure for the $\Lambda(1520)$ width in nuclei the following relative
observable--the double ratio: $R(^{A}X)/R(^{12}{\rm C})=
(\sigma_{pA}/A)/( \sigma_{p^{12}{\rm C}}/12)$,
i.e. the ratio of the nuclear total $\Lambda(1520)$ production cross section from $pA$ reactions divided by A to the same quantity on $^{12}$C.
But instead of this ratio, we consider the following analogous ratio
$R(^{A}X)/R(^{12}{\rm C})=
({\tilde \sigma}_{pA}(p_{\Lambda(1520)},0^{\circ})/A)/({\tilde \sigma}_{p^{12}{\rm C}}(p_{\Lambda(1520)},0^{\circ})/12)$, where
${\tilde \sigma}_{pA}(p_{\Lambda(1520)},0^{\circ})$ is the double differential cross section for the production
of $\Lambda(1520)$ hyperons with momentum $p_{\Lambda(1520)}$ at a lab angle of 0$^{\circ}$ in proton--nucleus collisions.
\begin{figure}[!h]
{\centering
\includegraphics[width=5.0cm]{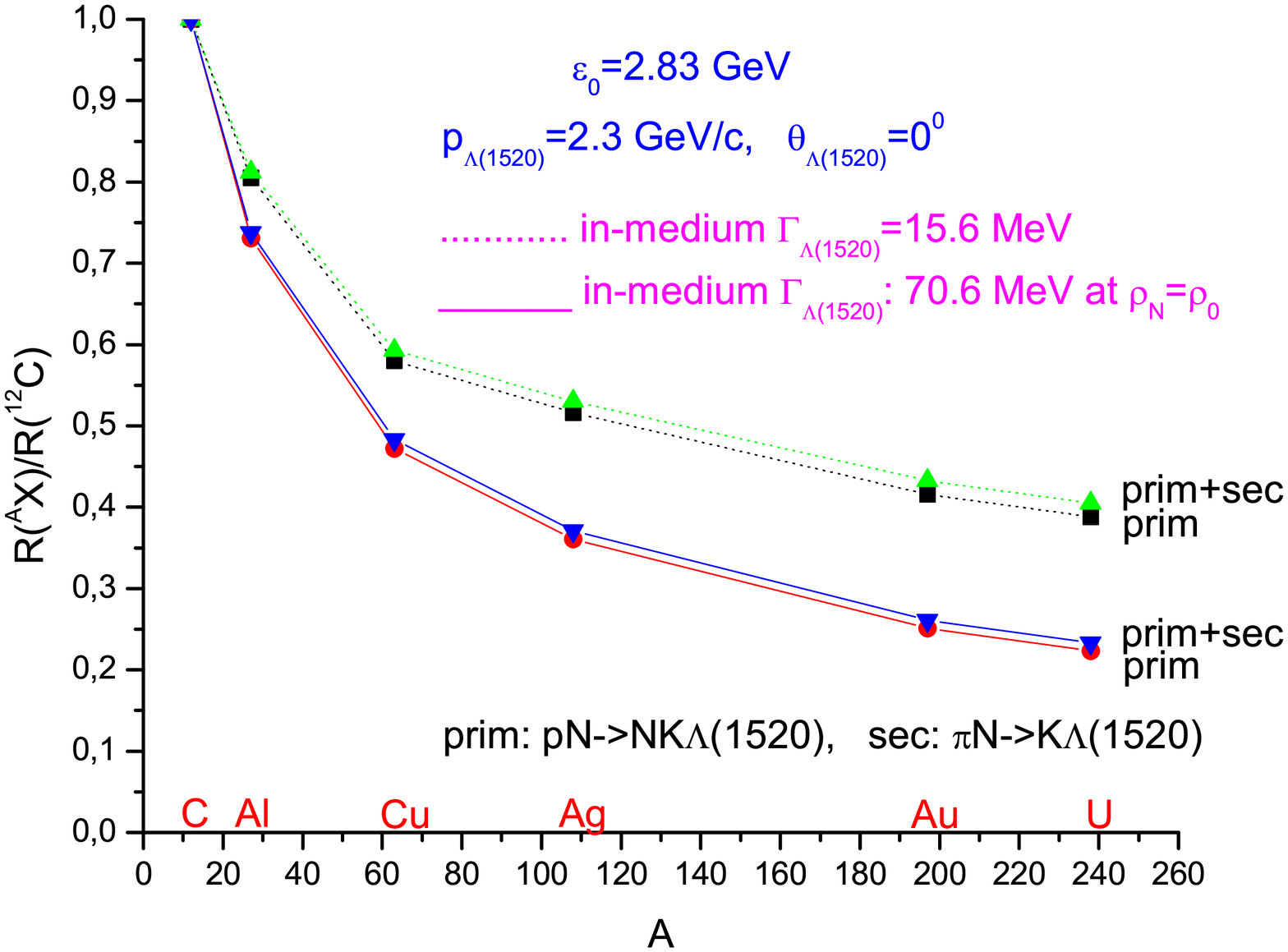}
\hfill
\includegraphics[width=5.0cm]{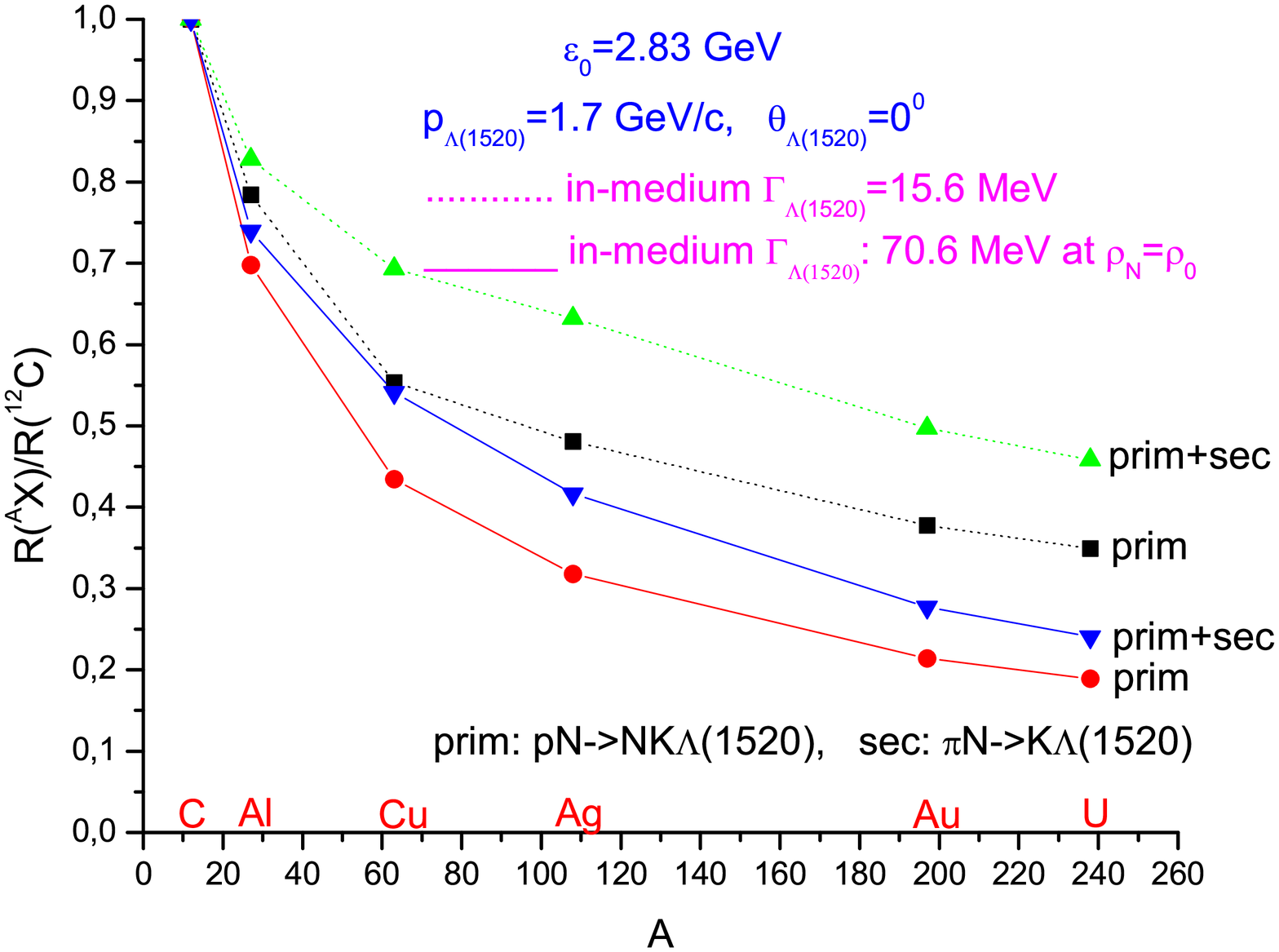}
\hfill
\includegraphics[width=5.0cm]{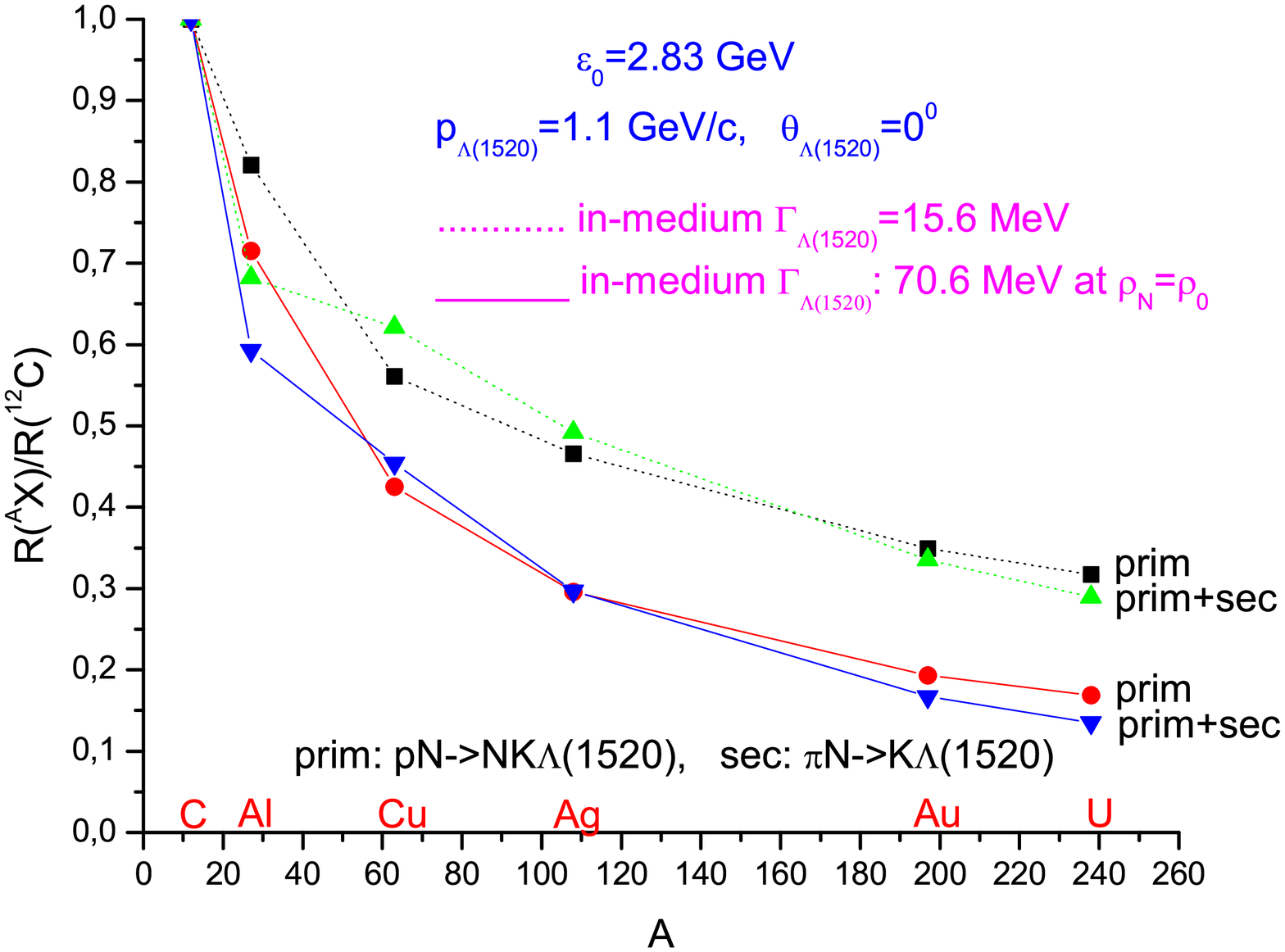}}
\vspace*{-2mm} \caption{Ratio
$R(^{A}X)/R(^{12}{\rm C})=
({\tilde \sigma}_{pA}(p_{\Lambda(1520)},0^{\circ})/A)/({\tilde \sigma}_{p^{12}{\rm C}}(p_{\Lambda(1520)},0^{\circ})/12)$
as a function of the nuclear mass number for initial energy of 2.83 GeV and for the
$\Lambda(1520)$ momenta of 2.3 GeV/c (left panel), 1.7 GeV/c (middle panel), 1.1 GeV/c (right panel) calculated
within the different scenarios for the $\Lambda(1520)$ hyperon production mechanism and for its in-medium width.
For notation see the text.}
\label{void}
\end{figure}

      Figure 8 shows the ratio
$R(^{A}X)/R(^{12}{\rm C})=
({\tilde \sigma}_{pA}(p_{\Lambda(1520)},0^{\circ})/A)/({\tilde \sigma}_{p^{12}{\rm C}}(p_{\Lambda(1520)},0^{\circ})/12)$
as a function of the nuclear mass number A calculated on the basis of equations (4), (30) for the one-step and
one- plus two-step $\Lambda(1520)$ creation mechanisms (corresponding lines with symbols 'prim' and 'prim+sec' by
them) for the projectile energy of 2.83 GeV as well as for the $\Lambda(1520)$ momenta of 2.3 GeV/c, 1.7 GeV/c,
1.1 GeV/c
and within
the considered two scenarios for the $\Lambda(1520)$ in-medium width: (i) free $\Lambda(1520)$ width (dotted lines),
(ii) in-medium $\Lambda(1520)$ width shown by solid curve in figure 4 (solid lines).
Looking at this figure, one can see that there are visible experimentally separated differences between the results
obtained by using different $\Lambda(1520)$ in-medium widths under consideration and the same assumptions concerning
the $\Lambda(1520)$ production mechanism (between corresponding dotted and solid lines). The secondary
pion-induced reaction channel ${\pi}N \to K\Lambda(1520)$ plays a minor role for all considered target nuclei
at $\Lambda(1520)$ momentum of 2.3 GeV/c
\footnote{Which is in line with our findings of figures 5, 6.}
.
Although this channel contributes dominantly to the absolute $\Lambda(1520)$ creation cross section at $\Lambda(1520)$
momentum of 1.1 GeV/c (see figures 5, 6), it manifests themselves insignificantly at this momentum in the
relative observable of interest for targets heavier than the Al target.
This is due to the fact that at $\Lambda(1520)$ momentum of 1.1 GeV/c the two-step to one-step $\Lambda(1520)$
production cross section ratio for $pA$ collisions is approximately equal to that for $p$C interactions. On the
other hand, at $\Lambda(1520)$ momentum of 1.7 GeV/c there are clear differences between calculations corresponding
to different suppositions about the $\Lambda(1520)$ creation mechanism and the same $\Lambda(1520)$ widths
in the medium (between dotted lines, and between solid lines in the middle panel of figure 8).
We may see, for example, that at this momentum for heavy nuclei the calculated ratio
$R(^{A}X)/R(^{12}{\rm C})$ can be of the order of 0.19 for the direct $\Lambda(1520)$ production mechanism
and 0.24 for the direct plus two-step $\Lambda(1520)$ creation mechanisms in the case when the absorption of
$\Lambda(1520)$ hyperons in nuclear matter was determined by their in-medium width shown by solid curve in
figure 4.
Since the main contribution to the nuclear total $\Lambda(1520)$ production cross section from direct mechanism
comes, as figure 6 shows, from the $\Lambda(1520)$ momenta $\sim$ 2.1--2.3 GeV/c as well as from small
$\Lambda(1520)$ production angles due to the kinematics, we can compare to a first approximation the above
ratio calculated with this width for the one-step $\Lambda(1520)$ creation mechanism at momentum of 2.3 GeV/c
with the corresponding ratio of the nuclear cross sections obtained in [20] at beam energy of 2.9 GeV
\footnote{Which is close to that of the present work.}
and for the nominal $\Lambda(1520)$ momentum-dependent in-medium width.
By looking at the solid line with symbol 'prim' by it in the left panel of figure 8
and at the solid curve in figure 5 (top panel) from [20], we obtain that our results exceed
the findings of [20] by factors of $\approx$ 1.1, 1.2 and 1.5 for target nuclei Al, Cu and U, respectively.
Such difference
in the A dependences for the $R$ ratios might be attributed in particular to the use in [20] the
momentum-dependent $\Lambda(1520)$ in-medium width.

Thus, we can conclude that the high precision
observation of the A dependence, like that considered above, can serve as an important tool to determine the $\Lambda(1520)$ width in nuclei and in the analysis of the observed dependence it is needed to account for in general the secondary
pion-induced $\Lambda(1520)$ production processes. To put more strong constraints on this width one needs to compare
the results of more reliable model calculations, based only on the direct $\Lambda(1520)$ production mechanism, with
the respective high-momentum data.
\begin{figure}[!h]
\begin{center}
\includegraphics[width=8.0cm]{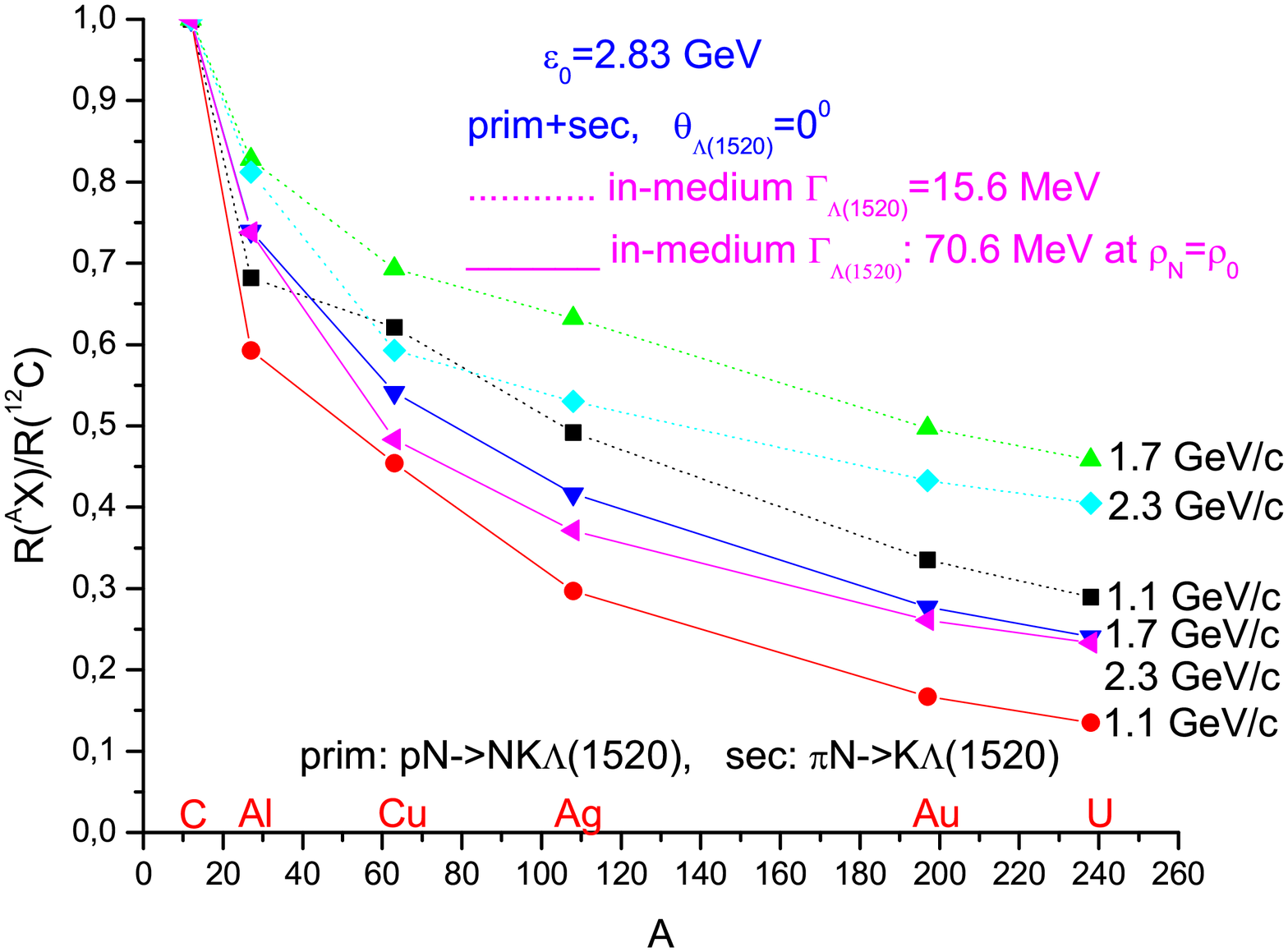}
\vspace*{-2mm} \caption{Ratio
$R(^{A}X)/R(^{12}{\rm C})=
({\tilde \sigma}_{pA}(p_{\Lambda(1520)},0^{\circ})/A)/({\tilde \sigma}_{p^{12}{\rm C}}(p_{\Lambda(1520)},0^{\circ})/12)$
as a function of the nuclear mass number for the one- plus two-step $\Lambda(1520)$ production mechanisms
calculated for incident energy of 2.83 GeV
within the different scenarios for the $\Lambda(1520)$ in-medium width and for the choice of the $\Lambda(1520)$
momentum. For notation see the text.}
\label{void}
\end{center}
\end{figure}

  Finally, in figure 9 we show together the results of our calculations, given before separately in figure 8,
  for the A dependence of our interest for the primary (1)--(3) plus secondary (29) $\Lambda(1520)$ production
  processes obtained for bombarding energy of 2.83 GeV by employing in them both the free $\Lambda(1520)$ hyperons
  width (dotted lines) and their in-medium width shown by solid curve in figure 4 (solid lines) as well as the
  considered options
  \footnote{Indicated by the respective symbols by the lines.}
  for the $\Lambda(1520)$ momentum to see also the sensitivity of the calculated A dependence to the $\Lambda(1520)$ momentum.
One can see that the differences between the calculations for the same $\Lambda(1520)$ width shown by solid curve in
figure 4 with adopting for the $\Lambda(1520)$ momentum two options: $p_{\Lambda(1520)}=1.7$ GeV/c and
$p_{\Lambda(1520)}=2.3$ GeV/c (between two upper solid curves) are insignificant, which means that this relative
observable depends weakly on the $\Lambda(1520)$ momentum in the high-momentum region
$1.7~{\rm {GeV/c}} \le p_{\Lambda(1520)} \le 2.3~{\rm {GeV/c}}$, where the cross section for $\Lambda(1520)$
production is larger.
The measurement of such weak dependence should, in particular, reflect the fact that the $\Lambda(1520)$ width in the medium is momentum independent.

   Thus, our results demonstrate that the measurements of the A and momentum dependences of
the absolute and relative cross sections for $\Lambda(1520)$ production in $pA$ reactions in the considered kinematics
and at above threshold beam energies will allow indeed to shed light on its in-medium properties.
They show also that to extract the definite information on the $\Lambda(1520)$
width in nuclear matter from the analysis of the measured such dependences it is needed to take into account in this analysis the secondary pion-induced $\Lambda(1520)$ production processes.

\section*{4. Conclusions}

\hspace{1.5cm} In this paper we have calculated the A and momentum dependences of the absolute and relative cross
sections for $\Lambda(1520)$ production at a lab angle of 0$^{\circ}$
from $pA$ reactions at 2.83 GeV beam energy by considering incoherent primary proton--nucleon and secondary pion--nucleon $\Lambda(1520)$
production processes in the framework of a nuclear spectral function approach [14, 15], which takes properly into account
the struck target nucleon momentum and removal energy distribution, calculated using the one-pion-exchange model
elementary cross sections for proton--nucleon
reaction channel as well as two different scenarios for the total $\Lambda(1520)$ width in the medium.
It was found that the secondary pion-induced reaction channel ${\pi}N \to K{\Lambda(1520)}$ contributes
distinctly to the low-momentum $\Lambda(1520)$ production in the chosen kinematics
and, hence, it has to be taken into consideration on close examination of these dependences with the aim to get information on the $\Lambda(1520)$ width in the nuclear matter. It was also shown that both
the A dependence of the relative $\Lambda(1520)$ hyperon production cross section and momentum dependence of the absolute $\Lambda(1520)$ hyperon yield from $pA$ collisions in the considered
kinematics and at incident energy of interest are appreciably sensitive to its in-medium width. This gives an
opportunity to determine the $\Lambda(1520)$ width in the nuclear medium at finite baryon densities experimentally.
\\
\\

{\bf Acknowledgments}\\

The author gratefully acknowledges M. Hartmann, Yu. Kiselev, V. Koptev, H. Str$\ddot{\rm o}$her for their interest in the work. This work is partly supported by the Russian Fund for Basic Research, Grant No.07-02-91565.

\end{document}